# Exploration of growth conditions of TaAs Weyl semimetal thin film by pulsed laser deposition[*]

Shien Li[1,2], Zefeng Lin [2,3], Wei Hu[2], Dayu Yan [2], Fucong Chen [2,3], Xinbo Bai[2,3], Beiyi Zhu[2], Jie Yuan [2,4], Youguo Shi [2,3,4], Kui Jin [2,3,4], Hongming Weng [2,3,4] and Haizhong Guo[1][†]

[1]*Key Laboratory of Materials Physics, Ministry of Education, School of Physics and Microelectronics, Zhengzhou University, Zhengzhou 450052, China*
[2] *Beijing National Laboratory for Condensed Matter Physics, Institute of Physics, Chinese Academy of Sciences, Beijing 100190, China*
[3] *School of Physical Sciences, University of Chinese Academy of Sciences, Beijing 100049, China*
[4] *Songshan Lake Materials Laboratory, Dongguan 523808, China*

TaAs, the first experimentally discovered Weyl semimetal material, has attracted a lot of attention due to its high carrier mobility, high anisotropy, nonmagnetic and strong interaction with light. These make it an ideal candidate for the study of Weyl fermions and the applications in quantum computation, thermoelectric devices, and photodetection. For further basic physics studies and potential applications, large-size and high-quality TaAs films are urgently needed. However, it is difficult to grow As-stoichiometry TaAs films due to the volatilization of As during the growth. To solve this problem, the TaAs films were attempted to grow on different substrates using targets with different As stoichiometric ratios by pulsed laser deposition (PLD). In this work, we have found that partial As ions of the GaAs substrate are likely to diffuse into the TaAs films during growth, which was preliminarily confirmed by the structural characterization, surface topography and composition analysis. As a result, the As content in the TaAs film is improved and the TaAs phase is achieved. Our work presents an effective method to fabricate the TaAs films by PLD, providing the possible use of the Weyl semimetal film for functional devices.

## 1. Introduction

Weyl semimetal (WSM) hosts Weyl fermions as quasiparticles in its low-energy excitation, with and only with Weyl nodes formed by crossing of two non-degenerate bands close enough to the chemical potential.[1-4] TaAs is the first recognized WSM without inversion centers, and also a kind of easier-to-grow and nonmagnetic material.[5] Intense interest has been attracted since the discovery of this topological material as it shows a series of peculiar phenomena, such as the observation of surface Fermi arcs,[6-

---

[*] Project supported by the National Key R&D Program of China (No. 2021YFA0718700), the National Natural Science Foundation of China (Nos. 12174347), and the Center for Materials Genome.
[†]Corresponding author. E-mail: hguo@zzu.edu.cn

[12] negative magnetoresistance phenomena due to chiral anomalies,[13] and Weyl nodes.[14]

In addition, TaAs shows high carrier mobility, distinct anisotropy in the electrical transport, and strong interaction with light,[15] [16] which therefore support the potential of TaAs in applications such as quantum computation, photodetection and thermoelectric devices.[17-19] For example, the photo detecting prototype based on the TaAs single crystal has been reported, and TaAs has the advantage of the wide temperature detection region due to its massless chiral Weyl nodes and fairly large absorptance over a broad spectral range.[19] The demonstrated generation of chiral ultrafast photocurrents of TaAs provides an unique opportunity for novel THz emission with polarization control.[18] Therefore, there exist many urgent requirements of large size and high quality TaAs films for the above mentioned potential applications.

At present, TaAs single crystal can be fabricated by chemical vapor transport.[20, 21] Recently, TaAs films were reported successfully grown via molecular beam epitaxy.[22, 23] To our knowledge, the TaAs thin film was never reported in literatures by pulsed laser deposition (PLD), which greatly limits its applications for the functional devices.[18, 24, 25] One serious challenge for the film preparation is that TaAs does not melt up to 1400 °C but decomposes before reaching its melting point. In addition, Ta-As binary system has complicated phase diagrams including $Ta_3As$, $Ta_2As$, $Ta_5As_4$, TaAs, and $TaAs_2$ phases, which leads to the difficulty to synthesize pure TaAs phase, especially in the TaAs film.[26, 27]

In this work, to explore the optimal growth condition of the TaAs films by PLD, Ta-As films were grown on various substrates using the TaAs targets with the different As contents, and their crystalline structures were characterized subsequently. The results reveal that the polycrystalline TaAs films can only be achieved on the GaAs substrates, which is likely to provide an effective As-rich atmosphere for TaAs, while the $Ta_5As_4$ phase is preferred to nucleate on other substrates without the As element. Above all, our study provides a practical approach to fabricate the designated films using the substrates of certain component-rich by PLD as well as other fabrication methods for the special volatile compounds and their related research.

## 2. Experimental details

First, we tried to fabricate the TaAs films on various small lattice mismatch substrates such as $SrTiO_3$, $0.7Pb(Mg_{1/3}Nb_{2/3})O_3$-$0.3PbTiO_3$ (PMN-PT), $MgF_2$ and MgO, using the $TaAs_{1+x}$ targets with the different As stoichiometric ratio such as TaAs, $TaAs_{1.1}$ and $TaAs_{1.2}$ by PLD method. Besides that, the TaAs films were also grown on the As compound substrate GaAs. All the samples were grown in a high vacuum of $10^{-7}$ Torr. The deposition temperature was set in the range from 650 °C to 750 °C. The fluence and repetition frequency of the KrF excimer laser ($\lambda = 248$ nm) were ~ 4 J/cm$^2$ and 5 Hz, respectively. The as-prepared powder of targets with nominal compositions of TaAs, $TaAs_{1.1}$ and $TaAs_{1.2}$ were prepared by solid-state reaction method, and the actual compositions were confirmed by Inductively Coupled Plasma (ICP) measurement. The targets were synthesized by compressing the powders into the pellets and then sintering at 450 °C for 24 hours in the vacuum furnace with $O_2$ and $H_2O$ levels both less than 10$^-$

² Torr. The structure characterization of the TaAs films was performed by Rigaku SmartLab X-ray diffractometer (XRD). The surface morphology was measured by Hitachi SU5000 scanning electron microscope (SEM). The composition was analyzed by energy-dispersive X-ray spectrometer (EDS).

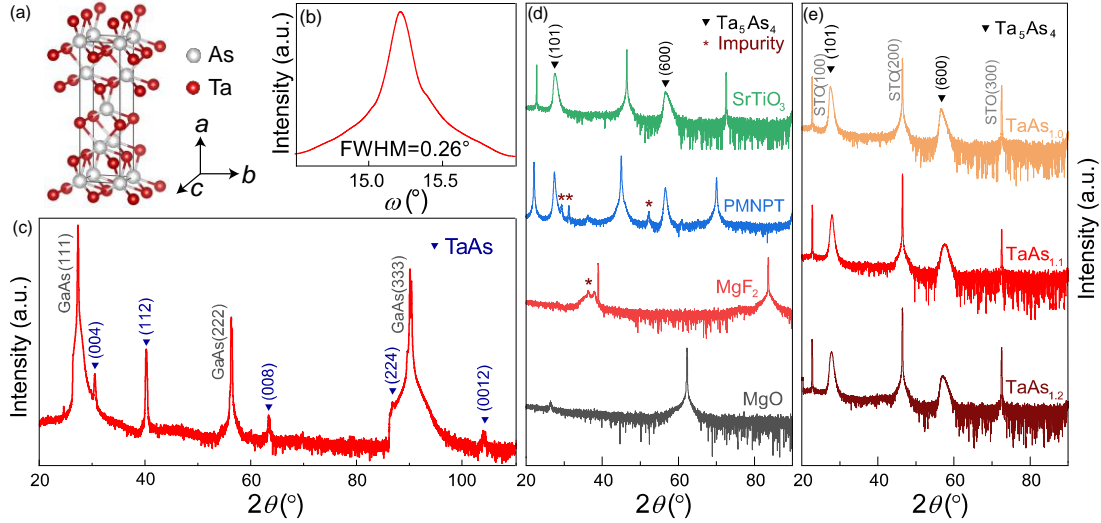

**Fig. 1.** (a) Body-centered tetragonal structure of TaAs with alternating stacked Ta and As layers. Crystal structure of TaAs does not have space inversion symmetry. (b) X-ray rocking curve of (004) reflections of the TaAs films, showing a full width half maximum of 0.26°. θ-2θ XRD scans of (c) the TaAs film on the GaAs substrate, (d) the $Ta_5As_4$ films on the different substrates, and (e) the $Ta_5As_4$ films using the TaAs targets with different As content.

## 3. Results and discussion
### 3.1. Film growth conditions

The crystal structure of TaAs is a body-centered tetragonal lattice system with lattice parameters $a$ = 3.437 Å and $c$ = 11.656 Å, and the space group is I41md (#109, $C_{4v}$) (Fig. 1(a)).[13, 28-30] First, we chose substrates with similar lattice structure or less mismatch to TaAs, such as $SrTiO_3$, PMN-PT, etc. The stoichiometric ratio of the TaAs target material is 1:1. Figure 1(d) shows the out-of-plane θ-2θ XRD scans of Ta-As films on $SrTiO_3$, PMN-PT, $MgF_2$ and MgO substrates. For Ta-As films on $SrTiO_3$ and PMN-PT substrates, the diffraction peaks at 2θ ≈ 27.87° and 57.58° were observed, indicating (101) and (600) reflections of $Ta_5As_4$, respectively, while the expected TaAs phase was not detected. According to the phase diagram,[26, 31] the appearance of $Ta_5As_4$ phase is due to the heavy loss of the As component during growth process.

In order to supplement As content in films, As-rich targets were used. Figure 1(e) shows the out-of-plane θ-2θ XRD scans of the Ta-As films on the $SrTiO_3$ substrates obtained by TaAs, $TaAs_{1.1}$, and $TaAs_{1.2}$ targets, respectively, but the expected TaAs phase was not obtained. During the preparation of the TaAs thin films, the target with the excess As component does not achieve the effect of the As supplementation. If the As component is supplemented by increasing the As ratio of the target, a large excess of the As component is required. However, the As component is easily sublimated, and there exists a confliction between the large stoichiometric ratio of As and the high compactness of the target, so the excess ratio of As in the target is limited.

The TaAs films were fabricated on GaAs substrates, which can provide As-rich environment. Figure 1(c) illustrates the $\theta$-$2\theta$ scans of a TaAs film on the GaAs (111) substrates, exhibiting (004), (008) and (112) diffraction peaks belonging to the TaAs phase. The rocking curve of the (004) peak shows a full width half maximum angle of 0.26° (Fig. 1(b)). Subsequently, we successfully fabricated the TaAs phase films with (004), (008) and (112) orientations on the GaAs (100) and (110) substrates using the same preparation conditions. Therefore, the lattice mismatch is probably not the key reason, while the As component in the substrates may play an important role on the nucleation of TaAs phase during the film deposition.

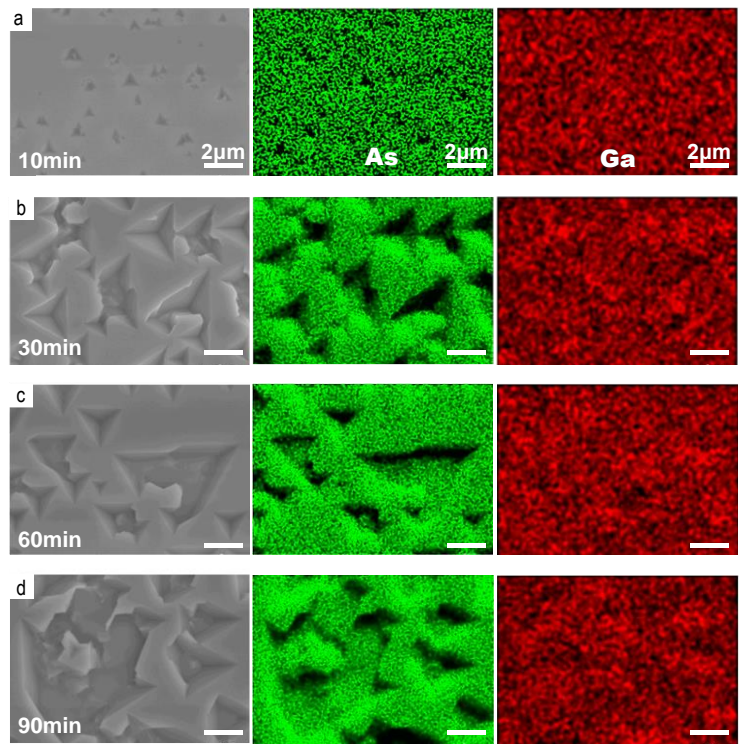

**Fig. 2.** SEM images and EDS elemental mapping of the TaAs films on the GaAs (111) substrates treated in 700 °C annealing for (a) 10 min, (b) 30 min, (c) 60 min, and (d) 90 min.

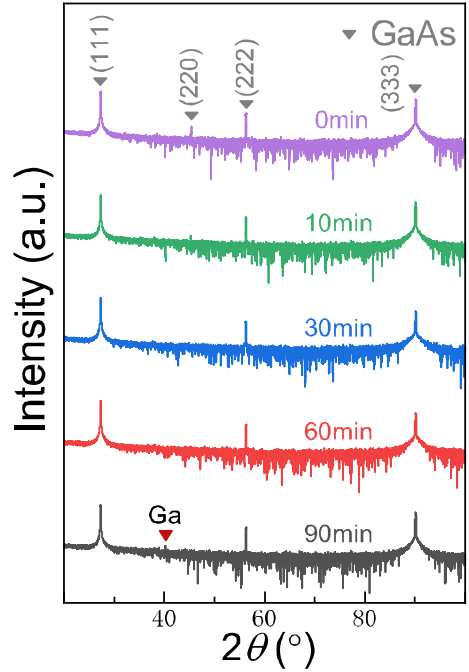

**Fig. 3.** The θ-2θ XRD scans of the TaAs films on the GaAs (111) substrates under different annealing times.

### 3.2. Evaporation of As from the substrate surface

The method of supplementing volatile compounds using the substrates containing the same volatile element is maybe a solution. It has been reported that the As content in the GaAs substrate decreases after annealing,[32] and a large number of As crystals appear on the surface of the substrate. In addition, some research groups carried out surface morphology characterization of the annealed GaAs substrate and found that the surface of the substrate was reconstructed.[33, 34] In order to explore the changes of the GaAs substrate during the growth of the TaAs films, we annealed the GaAs substrate at the growth temperature (700 °C) in a vacuum environment ($10^{-7}$ Torr). Then, the changes of the surface and composition of the substrates with different annealing time were investigated by SEM and EDS.

The SEM images of Figs. 2(a)-(d) show that the surface morphology of the substrate changes significantly after annealing for 10 min, 30 min, 60 min and 90 min under the same conditions. Under the same magnification ratio, the initial size of triangular cones on the surface is less than 1 μm, and it continues to expand to more than 5 μm with the increase of the annealing time. The EDS images in Fig. 2 show that with the increase of the annealing time, the area of As evaporation on the surface of the GaAs substrate gradually expands. By comparing the SEM and EDS images of the GaAs substrates at different annealing time, we demonstrate that the As content in the GaAs substrate evaporates from the surface of the substrate during the annealing process, which results in triangular cones on the surface of the substrate.

Figure 3 illustrates the out-of-plane θ-2θ XRD scans of the GaAs substrates annealed at different times. For GaAs substrates, the peaks at 2θ ≈ 45.37° represent the (220) reflection. It is found that with the increase of the annealing time, the (220) peak of the GaAs substrates gradually decreases and disappears, and the Ga peak appears at 90 min of annealing. The XRD data of the GaAs substrates with the different annealing times confirm our proposal that the As component in the GaAs substrate slowly evaporates under annealing conditions. It is proposed that the migration of the As component from the GaAs substrate to the films during the fabrication provides a new approach for the growth of the thin films containing volatile components.

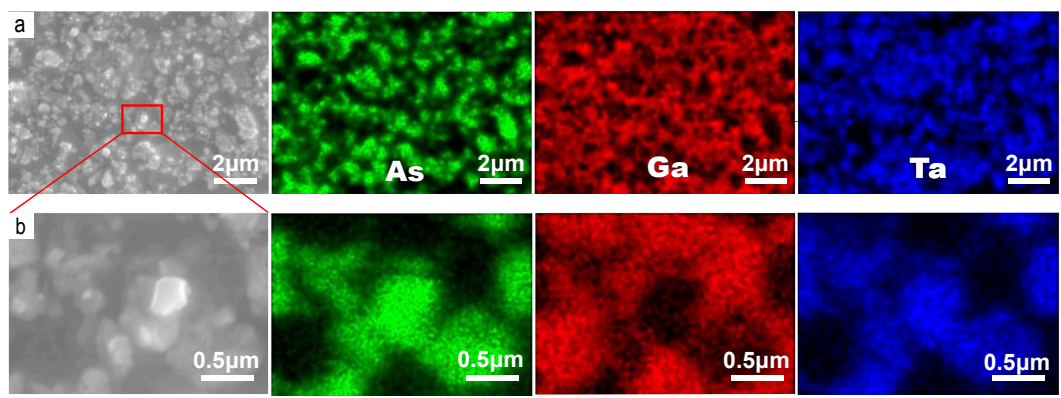

**Fig. 4.** (a) SEM images and EDS elemental mapping obtained in the TaAs films. (b) Enlarged SEM images and EDS elemental mapping of the red boxed area in (a).

### 3.3. Formation mechanism of the TaAs films

We have already found that the As component can evaporate from the GaAs substrate during the heating, and we try to figure out the formation mechanism of the TaAs thin films. As early as 1993, Han *et al.* reported that the ternary Ta-Ga-As phase could be formed at the interface between Ta and GaAs thin films due to the diffusion of As components at 850-950 °C.[26] Subsequently, Li *et al.* also found that the TaAs$_2$ phase annealed at 550 °C, and the TaAs, TaAs$_2$ and Cu$_3$Ga phases annealed at 600 °C were built in the Cu/Ta/GaAs multilayer films because of the percolation of As components.[35] Therefore, we realize that the As component in the GaAs substrate is unstable, which can penetrate to the neighboring Ta layer at high temperature.[36] At present, the TaAs film has only been reported to be prepared successfully on the GaAs substrates via molecular beam epitaxy.[22, 23]

In order to explore the TaAs film formation mechanism, we analyzed the TaAs thin film by SEM and EDX. The SEM images of Fig. 4(a) show that a large number of irregular crystal grains are randomly scattered on the surface of the TaAs film. In Fig. 4(b), we exhibit the detailed SEM images for the selected area in Fig. 4(a), and we can see that the film has multilayer structures of the crystal grains with 10 ~ 100 nm in diameters. The EDS images of Fig. 4(a) display the distribution of As and Ta components on the surface of the film, which is agreement with the distribution of Ga component. A detailed analysis of the selected region verifies the same conclusion (Figure 4(b)). Therefore, it can be determined that the crystal grain is of TaAs phase.

According to the discussion above, one can not achieve the TaAs film by PLD even with As-rich TaAs targets. By analyzing the SEM and EDS images of the GaAs substrates for various annealing times, we demonstrate that the As component can evaporate from the GaAs substrate during the annealing. Therefore, we successfully fabricate the TaAs films by PLD employing the GaAs substrate to supplement the volatile As component. The bottom-up method to replenish the volatile components from the substrate not only broadens the material types for the film preparation by PLD, but also inspires more ideas for all film fabrication techniques. Further studies are still required to improve the quality of the TaAs films by optimizing the surface treatment of the GaAs substrates.

## 4. Conclusion

In summary, we report a method of replenishing the volatile As component from the substrate, which enables us to successfully fabricate the TaAs film on the GaAs (111) substrate by PLD. The XRD data indicate that the TaAs film is of single phase because only (004) and (112) peaks are observed. By studying the SEM and EDS images of the GaAs substrates with different annealing times, we demonstrate that the GaAs substrate play an important role on the fabrication of the TaAs films, because the As component in the GaAs substrate can diffuse to the surface of the substrate during the annealing process. The method of supplementing volatile components through the substrate not only breaks through the traditional method of increasing the stoichiometric ratio of the volatile component in the target, but also provides the possibilities for the preparation of many special films with volatile components by PLD, as well as new inspiration for all film growth methods.

**Acknowledgements**
We would like to thank Z. Y. Huang, T. X. Liu, J. Xu, Z. Y. Zhao, Y. M. Zhang, Y. J. Shi, and Q. Li for helpful discussions.

**References**
[1] Wan X, Turner A M, Vishwanath A and Savrasov S Y 2011 *Phys. Rev. B* **83** 205101
[2] Xu G, Weng H, Wang Z, Dai X and Fang Z 2011 *Phys. Rev. Lett.* **107** 186806
[3] Bernevig B A 2015 *Nat. Phys.* **11** 698
[4] Weng H, Yu R, Hu X, Dai X and Fang Z 2015 *Adv. Phys.* **64** 227
[5] Weng H, Fang C, Fang Z, Bernevig B A and Dai X 2015 *Phys. Rev. X* **5** 011029
[6] Huang S, Xu S, Belopolski I, Lee C, Chang G, Wang B, Alidoust N, Bian G, Neupane M, Zhang C, Jia S, Bansil A, Lin H and Hasan M 2015 *Nat. Commun.* **6** 7373
[7] Xu S, Belopolski I, Alidoust N, *et al.* 2015 *Science* **349** 6248
[8] Lv B, Weng H, Fu B, Wang X, Miao H, Ma J, Richard P, Huang X, Zhao L, Chen G, Fang Z, Dai X, Qian T and Ding H 2015 *Phys. Rev. X* **5** 031013
[9] Xu S, Belopolski I, Sanchez D, *et al.* 2016 *Phys. Rev. Lett.* **116** 096801
[10] Lv B, Muff S, Qian T, Song Z, Nie S, Xu N, Richard P, Matt C, Plumb N, Zhao L, Chen G, Fang Z, Dai X, Dil J H, Mesot J, Shi M, Weng H and Ding H 2015 *Phys. Rev. Lett.* **115** 217601
[11] Yang L, Liu Z, Sun Y, Peng H, Yang H, Zhang T, Zhou B, Zhang Y, Guo Y, Rahn M, Prabhakaran D, Hussain Z, Mo S, Felser C, Yan B and Chen Y 2015 *Nat. Phys.*

**11** 879

[12] Liu Z, Yang L, Sun Y, Zhang T, Peng H, Yang H, Chen C, Zhang Y, Guo Y, Prabhakaran D, Schmidt M, Hussain Z, Mo S, Felser C, Yan B and Chen Y 2016 *Nat. Mater.* **15** 27

[13] Huang X, Zhao L, Long Y, Wang P, Chen D, Yang Z, Liang H, Xue M, Weng H, Fang Z, Dai X and Chen G 2015 *Phys. Rev. X* **5** 031023

[14] Lv B, Xu N, Weng H, Ma J, Richard P, Huang X, Zhao L, Chen G, Matt C, Bisti F, Strocov V, Mesot J, Fang Z, Dai X, Qian T, Shi M and Ding H 2015 *Nat. Phys.* **11** 724

[15] Zhang C, Yuan Z, Jiang Q, Tong B, Zhang C, Xie X C and Jia S 2017 *Phys. Rev. B* **95** 085202

[16] Osterhoudt G B, Diebel L K, Gray M J, Yang X, Stanco J, Huang X, Shen B, Ni N, Moll P J W, Ran Y and Burch K S 2019 *Nat Mater* **18** 471

[17] Peng B, Zhang H, Shao H, Lu H, Zhang D W and Zhu H 2016 *Nano Energy* **30** 225

[18] Gao Y, Kaushik S, Philip E J, Li Z, Qin Y, Liu Y, Zhang W, Su Y, Chen X, Weng H, Kharzeev D E, Liu M and Qi J 2020 *Nat. Commun.* **11** 720

[19] Chi S, Li Z, Xie Y, Zhao Y, Wang Z, Li L, Yu H, Wang G, Weng H, Zhang H and Wang J 2018 *Adv. Mater.* **30** 1801372

[20] Sankar R, Peramaiyan G, Muthuselvam I P, Xu S, Hasan M Z and Chou F C 2018 *J. Phys.: Condens. Matter* **30** 015803

[21] Li Z, Chen H, Jin S, Gan D, Wang W, Guo L and Chen X 2016 *Cryst. Growth Des.* **16** 1172

[22] Yanez W, Ou Y, Xiao R, Ghosh S, Dwivedi J, Steinebronn E, Richardella A, Mkhoyan K A and Samarth N 2022 arXiv: 2202.10656v1

[23] Sadowski J, Domagała J Z, Zajkowska W, Kret S, Seredyński B, Gryglas-Borysiewicz M, Ogorzałek Z, Bożek R and Pacuski W 2022 *Cryst. Growth Des.* **22** 6039

[24] Sun K, Sun S, Wei L, Guo C, Tian H, Chen G, Yang H and Li J 2017 *Chin. Phys. Lett.* **34** 117203

[25] Sirica N, Tobey R I, Zhao L, Chen G, Xu B, Yang R, Shen B, Yarotski D A, Bowlan P, Trugman S A, Zhu J, Dai Y, Azad A K, Ni N, Qiu X, Taylor A J and Prasankumar R P 2019 *Phys. Rev. Lett.* **122** 197401

[26] Han Q and Schmid-Fetzer R 1993 *Mater. Sci. Eng. B* **17** 147

[27] J.-O.Willerström 1984 *J. less-common met.* **99** 273

[28] Furuseth S, Selte K and Kjekshus A 1965 *Acta Chem. Scand.* **19** 95

[29] Murray J.J T J B, Calvert L.D, Wang Yu , GabeE.J , Despault J.G 1976 *J. less-common met.* **46** 311

[30] Zhang C, Xu S, Belopolski I, *et al.* 2016 *Nat. Commun.* **7** 10735

[31] Ghosh G and Materials Science International Team M 1994 *The Ta-As binary phase diagram at 1 bar: Datasheet from MSI Eureka in SpringerMaterials* (Stuttgart: MSI, Materials Science International Services GmbH）

[32] Campos C.E.M and Pizani P.S. 2002 *Appl. Surf. Sci.* **200** 111

[33] Zhang S and Northrup J E 1991 *Phys. Rev. Lett.* **67** 2339

[34] Biegelsen D K, Bringans R D, Northrup J E and Swartz L 1990 *Phys. Rev. B* **41** 5701
[35] Chen C, Chang L, Chang E, Chen S and Chang D 2000 *Appl. Phys. Lett.* **77** 3367
[36] Lahav A and Eizenberg M 1984 *Appl. Phys. Lett.* **45** 256